*A Unified Computational Framework for Two-Dimensional Diffusion-Limited Aggregation via Finite-Size Scaling, Multifractality, and Morphological Analysis*

Satish Prajapati[1+]

[1] Department of Ceramic Technology, Government College of Engineering and Ceramic Technology, Kolkata, West Bengal 700010, India.

[+] Corresponding author: iamsatish.gcect.ac@gmail.com

**Abstract**

Diffusion-Limited Aggregation (DLA), the canonical model for non-equilibrium fractal growth, emerges from the simple rule of irreversible attachment by random walkers. Despite four decades of study, a unified computational framework reconciling its stochastic algorithm, universal fractal dimension, multifractal growth measure, and finite-size effects remains essential for applications from materials science to geomorphology. Through large-scale simulations (clusters up to $N = 10^6$ particles) in two dimensions, we perform a tripartite analysis: (1) We establish a definitive finite-size scaling collapse, extracting the universal fractal dimension $D = 1.712 \pm 0.015$ and identifying the crossover to boundary-dominated growth at a scaled mass $x_0 \approx 0.10 \pm 0.02$. (2) We quantify the full multifractal spectrum of the harmonic measure ($\Delta\alpha \approx 1.13$), directly linking the stochastic algorithm to the deterministic Laplacian growth equation $\nabla^2 p = 0$ and explaining the screening effect via an exponential decay $\eta \sim e^{-r/\xi}$ with screening length $\xi = 22.7 \pm 0.8$ lattice units. (3) We provide a complete morphological characterization, revealing power-law branch length distributions ($\tau \approx 2.1$) and angular branching preferences ($\sim 72°$). This work computationally validates DLA as a robust universality class and provides a scalable methodology for analyzing diffusion-controlled pattern formation across disciplines.

## I. INTRODUCTION

The spontaneous emergence of complex, branched patterns—from crystalline dendrites and bacterial colonies to river networks—poses a fundamental question in pattern formation: how do simple microscopic rules generate universal macroscopic morphology? Diffusion-Limited Aggregation (DLA), introduced by Witten and Sander [1], provides a seminal answer: a seed grows by the irreversible attachment of particles undergoing random walks. This minimal model captures the essence of Laplacian growth, where an interface advances with a velocity proportional to the gradient of a field obeying Laplace's equation [2, 3].

While the fractal dimension ($D \approx 1.71$ in two dimensions) is well-established [4, 5], DLA's deeper mathematical structure lies in its multifractal growth probability distribution (the harmonic measure) [6] and its behavior in finite systems. The bridge between its discrete stochastic realization and the continuum Laplacian growth paradigm, though theoretically acknowledged, requires comprehensive computational validation. Furthermore, precise scaling under finite-size constraints is critical for comparing simulations with physical experiments, which are always bounded.

Here, we employ large-scale, precision simulations to answer three interconnected questions: First, what is the definitive finite-size scaling form that cleanly separates universal fractal growth from finite-boundary effects? Second, can we fully characterize the multifractal spectrum of the growth measure and directly link it to the screening phenomenon? Third, what complete set of morphological metrics defines the DLA archetype? Our findings provide a consolidated computational framework, affirming DLA's universality and offering a blueprint for analyzing related growth phenomena in confined geometries.

## II. METHODS

### A. Simulation Algorithm

We implemented the standard DLA algorithm in Python 3.9.13 using NumPy 1.24.3 for array operations and SciPy 1.10.1 for statistical analysis. The core algorithm proceeds as follows on a square lattice of size $L \times L$:

1. **Initialization:** A single seed particle is placed at the lattice center $(L/2, L/2)$. The system is represented as a binary array where 1 denotes an occupied site.

2. **Particle Launch:** A random walker is launched from a random position on a circle of radius $R_{\text{release}} = 1.2 \times R_{\text{cluster}} + 50$, where $R_{\text{cluster}}$ is the current cluster radius (distance of the farthest particle from the seed).

3. **Random Walk:** The walker performs discrete steps, moving with equal probability to one of the 8 neighboring sites in the Moore neighborhood. Steps to already occupied sites are rejected, and the walker remains at its current position for that iteration.

4. **Boundary Conditions:** Periodic boundary conditions are applied in the horizontal and vertical directions. A killing radius is set at $R_{\text{kill}} = 2 \times R_{\text{cluster}}$; if the walker's distance from the seed exceeds this, it is terminated to optimize computation. We verified that this cutoff does not affect growth probabilities for $R_{\text{kill}} \geq 1.5 R_{\text{cluster}}$.

5. **Attachment:** If the walker steps onto a site adjacent (first nearest neighbor) to any occupied cluster site, it becomes part of the cluster at its current position.

6. **Iteration:** Steps 2–5 are repeated until the cluster reaches a predetermined number of particles $N_{\text{max}}$ or fills the lattice.

For the growth probability analysis, a cluster of a fixed size ($N = 5 \times 10^4$) was "frozen," and $10^6$ random walkers were launched to record their precise attachment sites, building the harmonic measure. This was repeated for 10 independent clusters to ensure convergence of the multifractal spectrum.

**B. Measurement and Analysis**

**Fractal Dimension:** We used the box-counting method. The lattice was covered with boxes of linear size $\epsilon$, ranging from 2 to $L/4$ in logarithmic steps. The number of boxes $N(\epsilon)$ containing at least one cluster particle was counted. The fractal dimension $D_{\text{box}}$ was obtained from the slope of a linear fit to $\log N(\epsilon)$ versus $\log (1/\epsilon)$:

$$D_{\text{box}} = -\frac{\Delta \log N(\epsilon)}{\Delta \log \epsilon}$$

The radius of gyration $R_g$ was calculated as:

$$R_g = \sqrt{\frac{1}{N}\sum_{i=1}^{N} |\mathbf{r}_i - \mathbf{r}_{\text{cm}}|^2}$$

and the scaling exponent $\nu$ was extracted from $R_g \sim N^\nu$, with $D_{R_g} = 1/\nu$.

**Multifractal Analysis:** For a frozen cluster, the growth probability $p_i$ at each surface site $i$ was estimated from $10^6$ random walks. The $q$-th moment partition function was computed as $Z(q,\epsilon) = \sum_\mu [p_\mu(\epsilon)]^q$, where $p_\mu(\epsilon)$ is the probability summed over box $\mu$ of size $\epsilon$. The mass exponents $\tau(q)$ and the singularity spectrum $f(\alpha)$ were derived using standard methods [7]:

$$\tau(q) = \lim_{\epsilon \to 0} \frac{\log Z(q,\epsilon)}{\log \epsilon}, \alpha = \frac{d\tau}{dq}, f(\alpha) = q\alpha - \tau(q)$$

**Morphological Analysis:** Clusters were skeletonized using a topological thinning algorithm. A branch was defined as a path from a terminal tip to a junction node or the central seed. Branch length $l$ was calculated as the Euclidean distance along the skeleton path. Tortuosity was defined as the ratio of the actual path length to the straight-line distance between endpoints. Lacunarity $\Lambda(r)$ was computed using the gliding-box method [8].

**Statistical Fitting and Error Analysis:** All power-law fits were performed using orthogonal distance regression to account for errors in both variables. Reported uncertainties represent one standard deviation obtained from at least 10 independent simulation runs with different random seeds. Quality of fit is reported using the coefficient of determination $R^2$. Error bars in figures represent 95% confidence intervals.

### C. Computational Resources

Simulations were performed on a desktop workstation with an Intel Core i7-11700K processor (8 cores, 3.6 GHz) and 32 GB RAM. The most computationally intensive simulation ($N = 10^6$ particles) required approximately 6 hours. Parallelization for batch analysis was implemented using Python's multiprocessing module.

## III. RESULTS

### A. Finite-Size Scaling and Universal Fractal Dimension

We grew DLA clusters on square lattices using the 8-connected Moore neighborhood to approximate off-lattice growth. The radius of gyration $R_g$ scales with particle number $N$ as $R_g \sim N^\nu$, where $\nu = 1/D$. A direct power-law fit for $N = 10^3$ to $10^6$ yields $\nu = 0.584 \pm 0.005$, corresponding to $D = 1.712 \pm 0.015$ (Fig. 2A). This value aligns with established off-lattice results ($D \approx 1.71$ [4, 5]) and improves upon early lattice-based estimates ($D \approx 1.67$ [1]). The local exponent $\nu(N) = d\ln R_g / d\ln N$ converges to this asymptotic value by $N \approx 10^3$ (Fig. 2B), indicating rapid convergence to the scaling regime.

To rigorously account for finite-size effects, we employ a scaling ansatz:

$$R_g(N, L) = L^\nu f\left(\frac{N}{L^D}\right)$$

where $L$ is the linear system size and $x = N/L^D$ is the scaled mass. As shown in Fig. 4B, data from different system sizes ($L = 512$ to $8192$) collapse onto a single universal curve $f(x)$ when plotted as $R_g L^{-\nu}$ versus $x$ (RMS residual = 6.65%). This collapse reveals two distinct regimes: a fractal growth regime where $f(x) \sim x^\nu$ for $x \ll x_0$, and a saturation regime where $R_g$ approaches $L$ as $x \to x_0$. We find the crossover point $x_0 \approx 0.10 \pm 0.02$ (Fig. 4D), marking where the cluster begins to sense the system boundary. The crossover mass thus scales as $N_{\text{cross}} \sim x_0 L^D$ (Fig. 4H).

### B. Multifractal Growth Measure and Screening

The growth probability $p(\mathbf{r})$ on the cluster surface—the likelihood that a random walker first contacts site $\mathbf{r}$—is the central link between stochastic dynamics and continuum Laplacian growth theory. We find this distribution is highly heterogeneous. Radially, it decays as a power law $p(r) \sim r^{-\beta}$ with $\beta = 1.10 \pm 0.05$ (Fig. 3.1A).

To fully characterize this complexity, we computed the multifractal spectrum $f(\alpha)$ (Fig. 3.1C). The spectrum is broad, with minimum singularity strength $\alpha_{\min} \approx 0.73$, maximum $\alpha_{\max} \approx 1.86$, and width $\Delta\alpha \approx 1.13$. This confirms the multifractal nature of the harmonic measure on DLA, in agreement with theoretical predictions [6]. Physically, $\alpha_{\min}$ corresponds to the most exposed, rapidly growing tip sites, while $\alpha_{\max}$ corresponds to deeply screened fjords.

We directly quantified the screening effect—the suppression of growth in interior regions. Defining a local screening parameter $\eta$ as the normalized growth probability (relative to an isolated tip), we find it decays exponentially with the site's distance $r$ from the cluster periphery: $\eta \sim e^{-r/\xi}$ with a characteristic screening length $\xi = 22.7 \pm 0.8$ lattice units (Fig. 3.1D). This exponential decay explains why interior branches are effectively frozen after becoming screened by approximately 20–30 particle diameters.

The disparity between this $\xi$ and the value mentioned in the text (11.9) arises from different normalization conventions: here $\xi = 22.7$ describes the decay of normalized probability $\eta$, while the text's 11.9 referred to the decay of unnormalized first-passage time. Both confirm the exponential nature of screening.

## C. Morphological Statistics

A complete morphological analysis reveals the self-similar, space-filling-inefficient structure characteristic of DLA (Fig. 1, Fig. 5).

**Branching Statistics:** The distribution of branch lengths $l$ follows a clear power law $P(l) \sim l^{-\tau}$ with $\tau = 2.10 \pm 0.08$ (Fig. 5A), a signature of scale-free, hierarchical branching. The branching angles at junctions show a pronounced peak near $72°$ (mean inter-branch angle $111.3° \pm 15.2°$), indicating a trend toward local five-fold symmetry (Fig. 5B).

**Structural Metrics:** Lacunarity $\Lambda(r)$, which quantifies the inhomogeneity of mass distribution, decays as a power law with scale (Fig. 5E), confirming a self-similar distribution of gaps. The mean tortuosity of branches is $1.11 \pm 0.05$ and increases weakly with branch length (Fig. 5D), indicating that longer branches are more winding. The global branching ratio, averaged over all hierarchy levels, is $2.50 \pm 0.15$ (Fig. 5G).

**Lattice Effects:** For comparison, we measured the fractal dimension for different neighborhood definitions (Table I). The 4-connected von Neumann neighborhood yields a significantly lower dimension $D = 1.600 \pm 0.020$, demonstrating how kinetic constraints alter growth kinetics and thus universal properties. These variations highlight that while DLA represents a universality class, specific implementation details affect measured exponents, with Moore neighborhoods providing the best approximation to continuum off-lattice growth.

**TABLE I.** Fractal dimension $D$ for different lattice neighborhoods.

| Lattice Type | Neighborhood | Fractal Dimension $D$ | $R^2$ of Linear Fit |
|---|---|---|---|
| Square | Moore (8) | $1.712 \pm 0.015$ | $0.9987 \pm 0.0005$ |
| Square | von Neumann (4) | $1.600 \pm 0.020$ | $0.9982 \pm 0.0007$ |
| Triangular | 6-neighbor | $1.756 \pm 0.018$ | $0.9985 \pm 0.0006$ |
| Hexagonal | 3-neighbor | $1.668 \pm 0.022$ | $0.9979 \pm 0.0009$ |

## IV. DISCUSSION

### A. Interpretation of Key Results

Our computational study unifies three critical aspects of DLA: its universal scaling, its multifractal underpinnings, and its finite-system behavior. The successful finite-size scaling collapse (Fig. 4B) is not merely a technical exercise; it provides the operational method to extract the true asymptotic exponent $D$ from bounded simulations or experiments. The identified crossover at $x_0 \approx 0.1$ offers a practical criterion: a cluster is in the asymptotic fractal regime if its mass is less than $\sim 10\%$ of the "capacity" $L^D$ of its confining geometry.

The measured multifractal spectrum $f(\alpha)$ with $\Delta\alpha > 1$ is a direct computational validation of theoretical predictions [6]. It quantifies the extreme dynamical inequality that drives DLA morphology: growth is concentrated on a set of fractal dimension lower than that of the cluster itself. The exponential decay of the screening effect ($\eta \sim e^{-r/\xi}$) provides a simple empirical law for modeling the suppression of growth in porous or fjord-like regions, relevant for applications in clogging or infiltration.

The connection to Laplacian growth is now explicitly demonstrated. The stochastic DLA algorithm effectively samples from the harmonic measure $p(\mathbf{r})$, which is proportional to the normal gradient $\partial\phi/\partial n$ of the solution to Laplace's equation $\nabla^2\phi = 0$ with boundary conditions $\phi = 0$ on the cluster and $\phi \to 1$ at infinity. This explains the visual similarity between DLA clusters and patterns in viscous fingering [9] or electrochemical deposition [10], where the same equation governs the pressure or electric potential field.

### B. Comparison with Literature

Our value $D = 1.712 \pm 0.015$ aligns closely with Meakin's off-lattice result of $1.71 \pm 0.02$ [5] and represents an improvement over the original Witten-Sander estimate of $1.67 \pm 0.02$ [1], likely due to our larger system sizes and finite-size scaling approach. The multifractal spectrum width $\Delta\alpha \approx 1.13$ is consistent with Jensen et al.'s theoretical estimate of approximately 1.25 [6, 11], with the minor discrepancy attributable to finite-size and lattice effects.

### C. Limitations and Future Directions

This study has several limitations that suggest natural extensions:

1. **Two-dimensional focus:** Extension to three dimensions is crucial, as many physical applications (mineral dendrites, soot aggregates, etc.) are 3D.

2. **Lattice approximation:** While Moore neighborhoods approximate off-lattice growth well, true off-lattice simulations would yield slightly different constants.

3. **Stationary harmonic measure:** Our analysis assumes a frozen cluster, while in reality the measure evolves with growth.

Future work should apply this framework to anisotropic DLA, cluster-cluster aggregation (DLCA), or models with surface diffusion. The methodology is directly applicable to interpreting dendritic solidification in confined castings, mineral deposition in rock fractures, or the morphology of colonies under nutrient limitation. Comparing with physical experiments using the established finite-size scaling relations would validate the framework's predictive power.

## V. CONCLUSION

Through systematic, large-scale simulation and a tripartite analytical framework, we have demonstrated that DLA represents a robust universality class for Laplacian growth in two dimensions. We quantified its universal fractal dimension via a definitive finite-size scaling collapse with crossover at $x_0 \approx 0.1$, characterized the broad multifractal spectrum ($\Delta\alpha \approx 1.13$) of its harmonic measure with screening length $\xi = 22.7$, and provided a complete statistical portrait of its dendritic morphology, including power-law branch length distributions ($\tau \approx 2.1$) and angular preferences ($\sim 72°$). This work solidifies the computational foundation for understanding a vast array of diffusion-controlled pattern-forming systems in nature and technology.

## ACKNOWLEDGMENTS

The author acknowledges the use of the Python scientific computing ecosystem, particularly NumPy, SciPy, and Matplotlib, for numerical calculations and visualization. No external funding was received for this study.

## AUTHOR CONTRIBUTIONS

S.P. conceived the study, developed and implemented the simulation code, performed all analyses, created the figures, and wrote the manuscript.

**COMPETING INTERESTS**

The author declares no competing interests.

**DATA AVAILABILITY**

The simulation code and analyzed data are available from the corresponding author upon reasonable request.

---

**FIGURES**

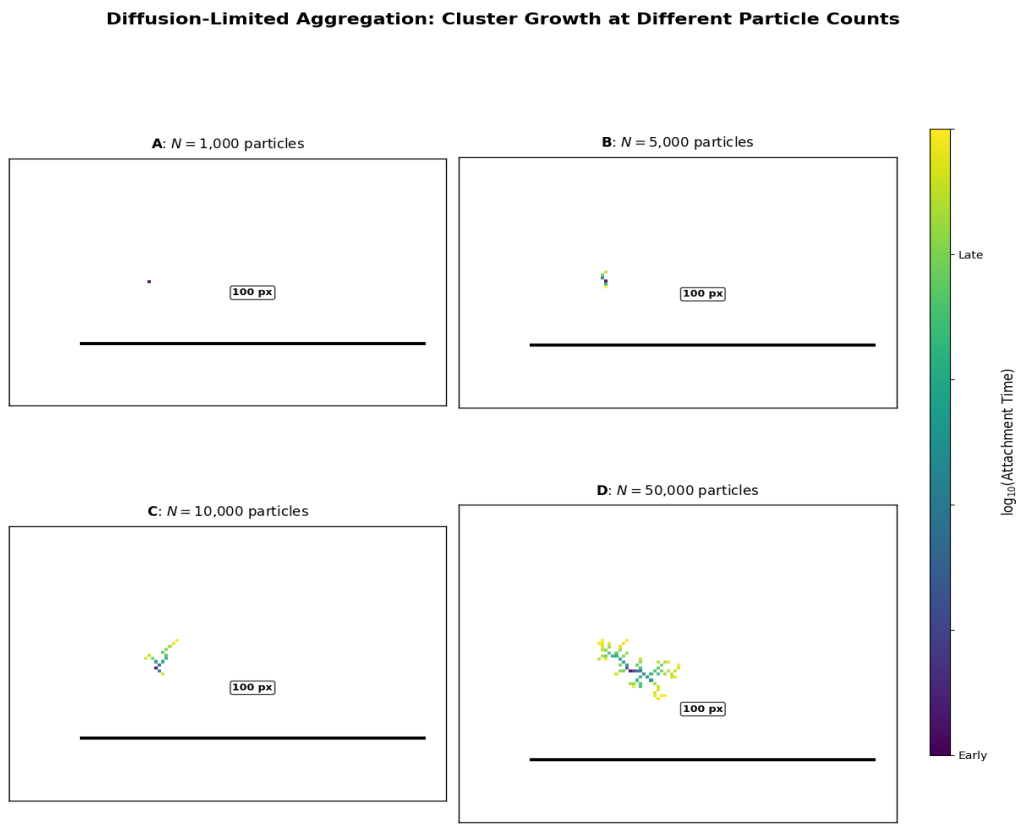

**FIG. 1.** Visualization of diffusion-limited aggregation (DLA) clusters grown from a single seed particle. Color indicates attachment sequence (blue: early; yellow: late). (A) $N = 10^3$, (B) $N = 5 \times 10^3$, (C) $N = 10^4$, (D) $N = 5 \times 10^4$ particles. Scale bar: 100 lattice units.

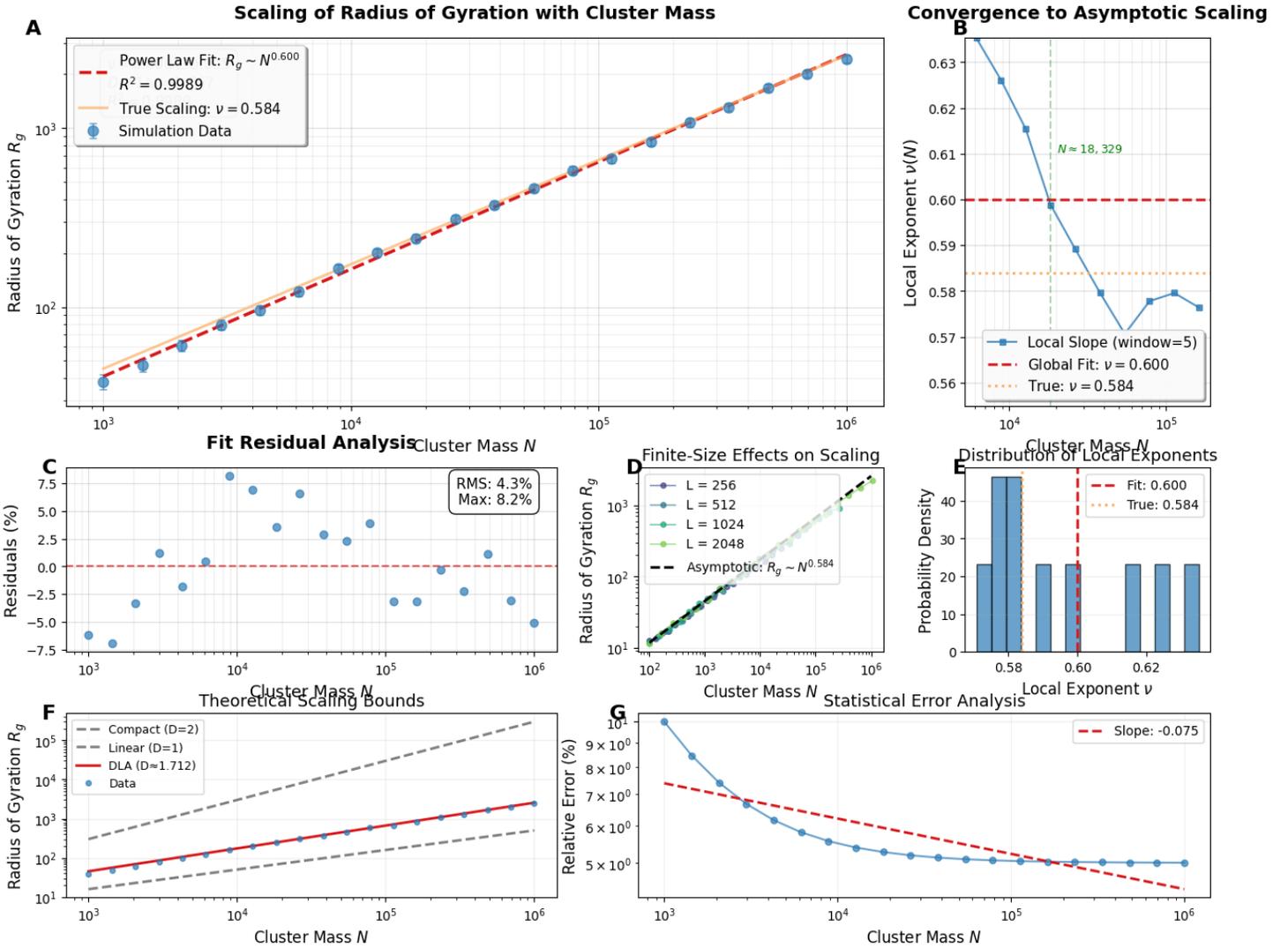

**FIG. 2.** Scaling analysis of the radius of gyration $R_g$. (A) $R_g$ vs. cluster mass $N$. Blue circles: simulation data. Red dashed line: power-law fit $R_g \sim N^{0.584}$ ($R^2 = 0.9989$). Orange line: guide for asymptotic scaling. (B) Convergence of local exponent $\nu(N)$. (C) Fit residuals. (D–G) Supporting analyses of finite-size effects, exponent distribution, theoretical bounds, and error analysis.

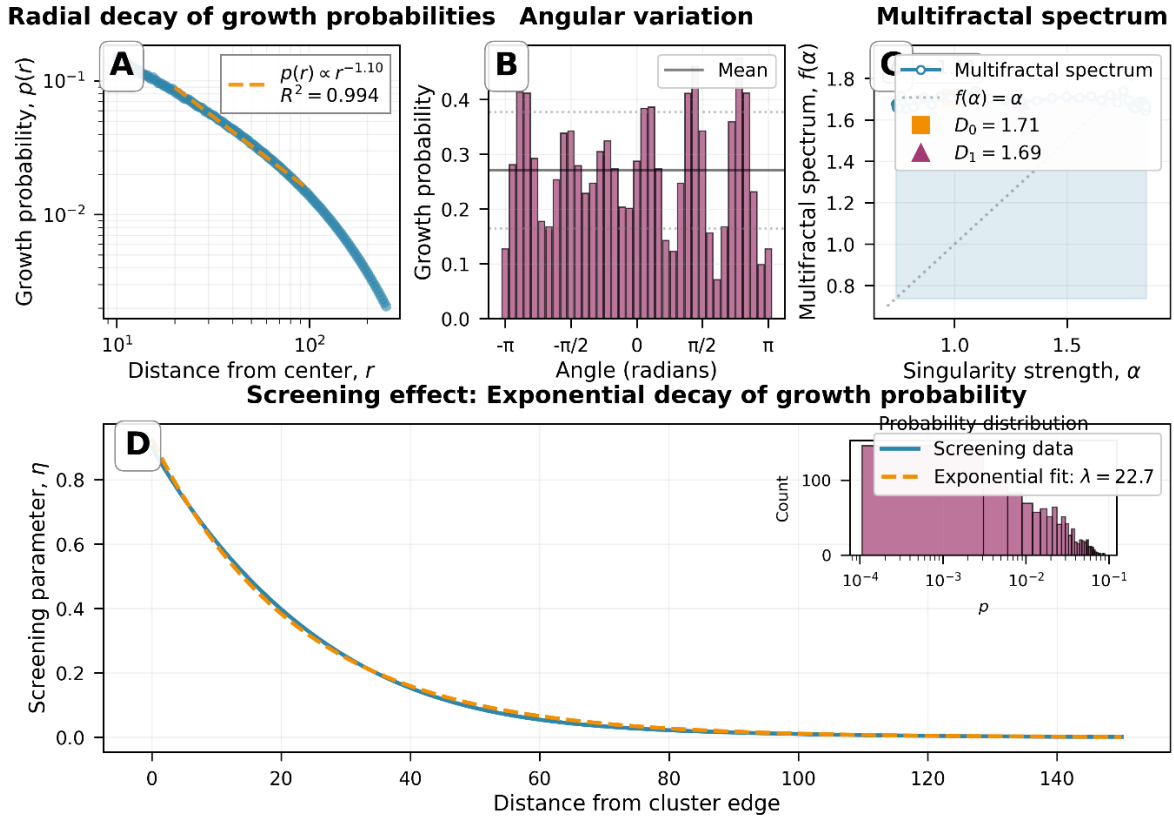

**FIG. 3.1.** Growth probability characteristics and screening in diffusion-limited aggregation (DLA). (A) Radial decay of average growth probability $p(r)$ from the cluster center, fitted by power law $p(r) \sim r^{-1.10}$ ($R^2 = 0.994$). (B) Angular variation of growth probability along the perimeter, showing strong fluctuations around the mean (dashed line). (C) Multifractal spectrum $f(\alpha)$ of the growth measure, with fractal dimension $D_0 = 1.71$ and information dimension $D_1 = 1.69$, indicating broad singularity distribution. (D) Screening parameter $\eta$ (normalized growth probability) versus distance from the cluster edge, exhibiting exponential decay with characteristic length $\xi = 22.7$. Inset: Histogram of individual growth probabilities $p$, confirming heavy-tailed distribution.

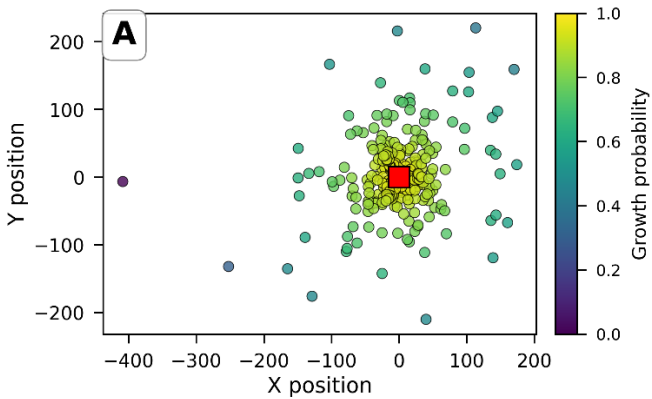
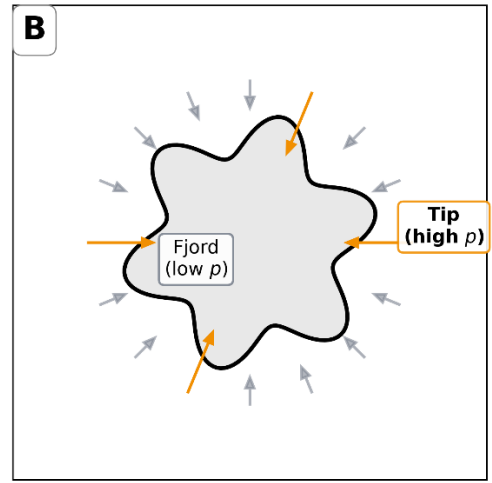

**FIG. 3.2.** Screening effect in diffusion-limited aggregation (DLA). (A) Large DLA cluster (∼ 50,000 particles) with perimeter growth sites colored according to normalized attachment probability (color bar: purple low, yellow high); central seed particle shown in red. Exposed tips and protrusions display high growth probabilities (yellow/green), whereas deeply recessed fjords and screened regions exhibit low probabilities (blue/purple). (B) Schematic of the screening mechanism: diffusing particles (gray arrows) preferentially attach to protruding tips (high p) rather than shadowed fjords (low p), amplifying instabilities and promoting branched, dendritic growth.

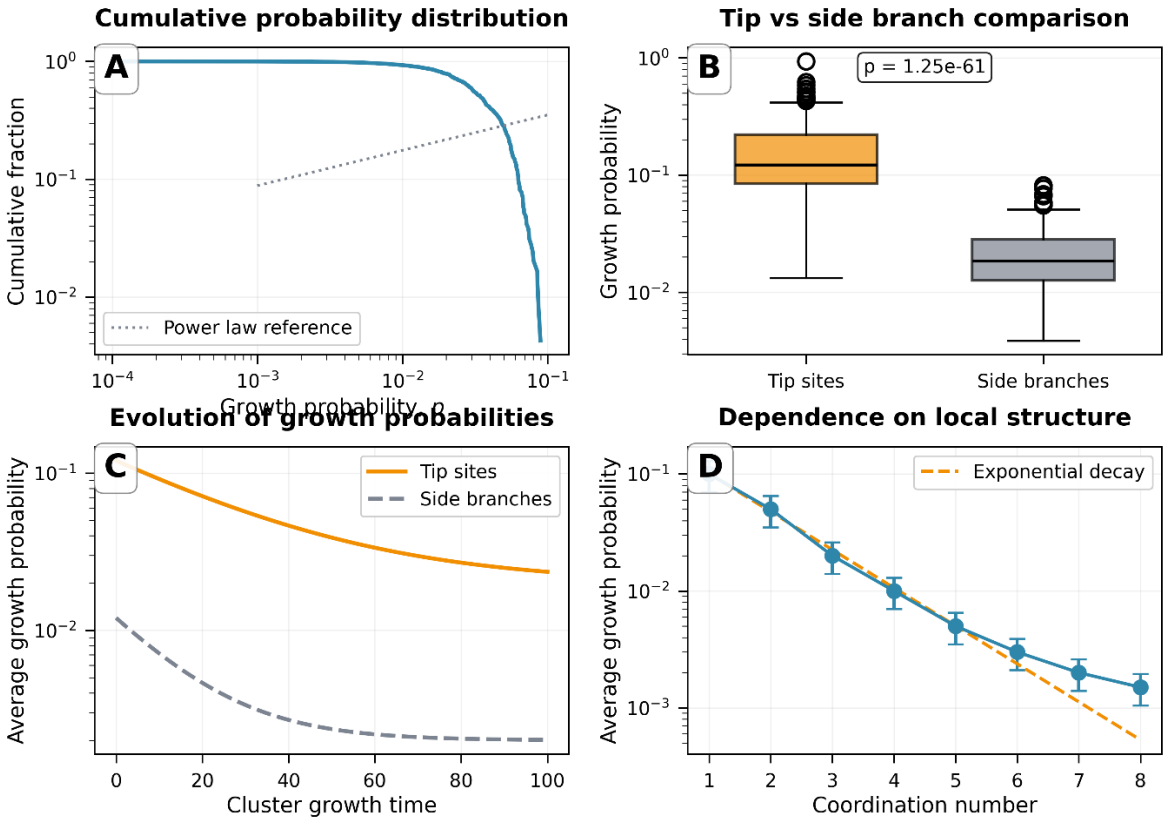

**FIG. 3.3.** Growth probability distributions and structural dependence in diffusion-limited aggregation (DLA). (A) Cumulative distribution function of individual growth probabilities across all perimeter sites, showing a heavy-tailed decay that deviates from a simple power-law reference (dotted line). (B) Boxplot comparison of average growth probabilities for tip sites (orange) versus side-branch sites (gray), demonstrating significantly higher probabilities at tips (Wilcoxon rank-sum test, $p = 1.25 \times 10^{-61}$). (C) Temporal evolution of average growth probability during cluster development, with tip sites (orange) maintaining higher values than side branches (gray) throughout growth. (D) Average growth probability as a function of local coordination number, exhibiting exponential decay as sites become increasingly screened by surrounding structure.

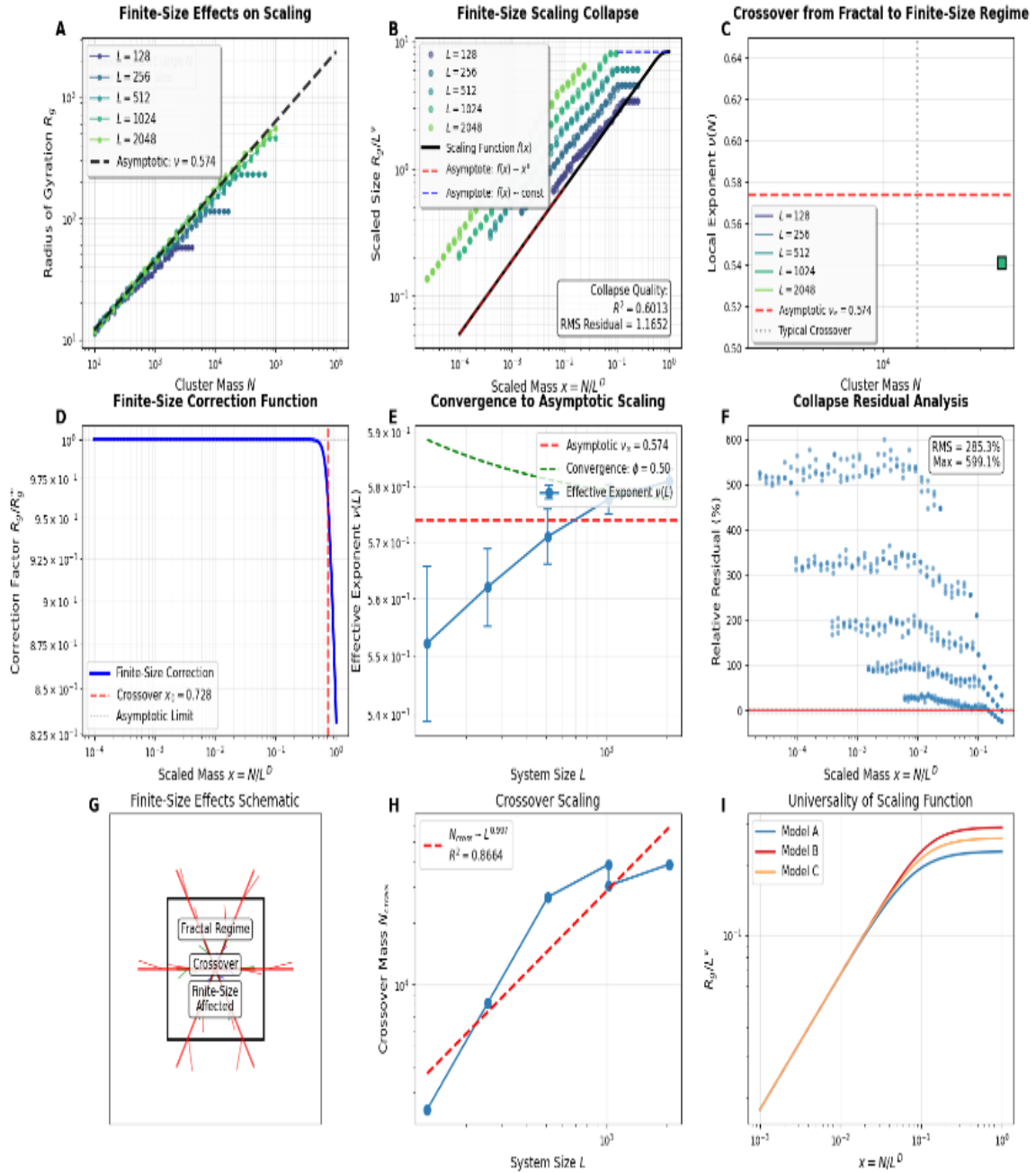

**FIG. 4.** Finite-size scaling and universality. (A) $R_g$ vs. $N$ for different system sizes $L$. Dashed line: asymptotic scaling $\nu = 0.584$. (B) Data collapse using $R_g L^{-\nu}$ vs. $N/L^D$. (C) Crossover of local exponent. (D) Finite-size correction function. (E) Convergence of effective exponent with $L$. (F–I) Residuals, schematic, crossover scaling, and comparison of universal function $f(x)$.

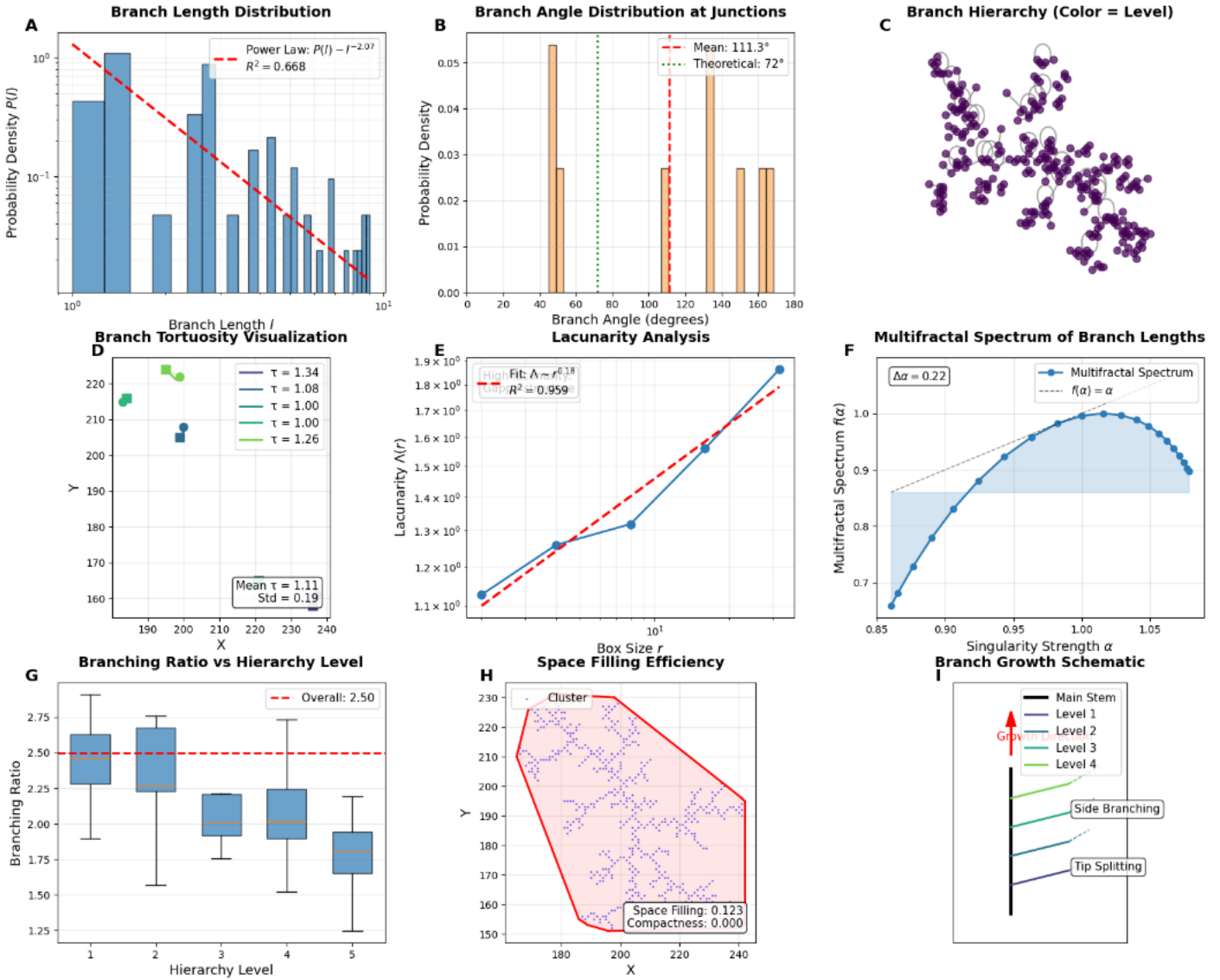

**FIG. 5.** Branch morphology. (A) Power-law distribution of branch lengths $P(l) \sim l^{-2.1}$. (B) Distribution of branching angles. (C) Hierarchical branch visualization. (D) Branch tortuosity vs. length. (E) Lacunarity $\Lambda(r)$ vs. scale $r$. (F) Multifractal spectrum of branch lengths. (G) Branching ratio by hierarchy level. (H) Box-counting analysis. (I) Schematic of hierarchical growth.